\documentclass[%
 reprint,
showpacs,%preprintnumbers,
 amsmath,amssymb,
 aps,
] {revtex4-1}

\usepackage{graphicx}% Include figure files
\usepackage{subfig}
\usepackage{float}
\floatstyle{boxed}
\graphicspath{{D:\Me_Paper\manuscript_jcp}}
\usepackage{dcolumn}% Align table columns on decimal point
\usepackage{bm}% bold math
\usepackage{xcolor}

\newcommand{\ti}{\textsuperscript}

\begin{document}
\title{Signatures of nonlinear magnetoelectricity in second harmonic spectra of SU(2) symmetry broken quantum many-body systems}
\author{Abhiroop Lahiri}
\email{lahiri.abhiroop@gmail.com}
\author{Swapan K. Pati}
\affiliation{ Theoretical Sciences Unit, Jawaharlal Nehru Centre for 
 Advanced Scientific Research, Jakkur, Bangalore-560064, India}
\date{\today}
\begin{abstract}
Quantum mechanical perturbative expressions for second order dynamical magnetoelectric (ME) susceptibilities have been derived and calculated for a small molecular system using the Hubbard Hamiltonian with SU(2) symmetry breaking in the form of spin-orbit coupling (SOC) or spin-phonon coupling. These susceptibilities will have signatures in second harmonic generation spectra. We show that SU(2) symmetry breaking is the key to generate these susceptibilities. We have calculated these ME coefficients by solving the Hamiltonian for low lying excited states using Lanczos method.
Varying the Hubbard term along with SOC strength, we find spin and charge
and both spin-charge dominated spectra of dynamical ME coefficients. We have shown that intensities of the peaks in the spectra are highest when the
magnitudes of Hubbard term and SOC coupling term are in similar range.
\end{abstract}
\maketitle
\section{INTRODUCTION}
  The study of Magnetoelectric(ME) effect in materials has gained a huge 
interest due to their potential applications in sensors \cite{1},
ME RAM \cite{2,3,4}  and other spintronic devices.
 The ME effect is observed in materials where there is a significant
coupling between the charge and the spin degrees of freedom, . This coupling gives rise to interesting linear as well as nonlinear responses in the presence 
of an electromagnetic field and are manifested in the measurement of nonlinear optical 
susceptibilities. Several mechanisms of this ME effect have been proposed. 
The most prominent one is due to exchange striction \cite{5} which not 
only explains ME effect in some multiferroic oxides\cite{6,7}, but also organic 
molecular solids \cite{8}. Others are due to inverse Dzyaloshinskii-Moriya 
interaction or spin current mechanism \cite{9,10,11} in spiral magnets and 
spin-dependent metal-ligand hybridization \cite{12,13,14,15} which involve 
spin-orbit coupling. In recent years, electrical control of magnetism have also been discovered in layered 2D materials such as bilayer MoS$_2$\cite{mos2}, CrI$_3$\cite{cri3}, etc.
  There has also been many first principles studies of ME materials\cite{spaldin,1st_principle_1,1st_principle_2}
Still it is essential to know the role of electronic correlations in the manifestation of this phenomenon.
In this article, we have theoretically calculated the second order nonlinear 
ME susceptibilities in small molecular systems using weak incident light as 
perturbation using a local many-body Hamiltonian. These susceptibilities will appear as small peaks in the second harmonic generation spectrum of the materials.
\section{NONLINEAR MAGNETOELECTRIC SUSCEPTIBILITIES FROM PERTURBATION THEORY}
Following the phenomenological theory of Landau \cite{a}, the second order ME contribution to the polarization ($P$) and 
magnetization ($M$) at the direction $\alpha$, in the presence of electromagnetic 
field ($E$,$B$), can be written as

\vspace{-5.0pt}
  
\begin{equation}\label{eq:polar}
  \tilde{P}_{\alpha}(\bm{E},\bm{B}) = -\frac{\partial F}{\partial E_{\alpha}} \big|_{\bm{B}}
                                      = \tilde{\chi}^{(2)}_{eem}\bm{EB}
\end{equation}
and  
\begin{equation}\label{eq:mag}
  \tilde{M}_{\alpha}(\bm{E},\bm{B}) = -\frac{\partial F}{\partial B_{\alpha}} \big|_{\bm{E}}
   = \tilde{\chi}^{(2)}_{emm}\bm{EB}
\end{equation}
where $F$ is the free energy of the system. $\tilde{\chi}^{(2)}_{eem}$ and $\tilde{\chi}^{(2)}_{emm}$ are two types of second order ME susceptibilities for the instances when the free energy F is proportional to $\bm{EEB}$ and $\bm{EBB}$. In this article, we have focussed on $\tilde{\chi}^{(2)}_{eem}$, although the inferences of the results will be valid for $\tilde{\chi}^{(2)}_{emm}$ also.
Light has been used as a 'probe' to find the ME coefficients. The electric and magnetic 
fields of light couple with the electric dipole  moment and the spin, thus manifesting 
the non-linear ME effects in the polarization or magnetization.
To derive the expressions for ME susceptibilities, we consider a general Hamiltonian 
\begin{equation}
	\hat{H} = \hat{H_0} + \hat{H_1}
\end{equation}
where $H_0$ is any arbitrary Hamiltonian and $H_1$ is the perturbation.
$H_1$  can be written as
\begin{equation}\label{eq:perturb}
\begin{split}
H_1& = -\displaystyle\sum_{i}{n_i\bm{E}.\bm{\hat r_i}} - \displaystyle\sum_{i}{\bm{B}.\bm{\hat S_i}}\\
& = -\bm{\mu .E} - \bm{\nu .B}
\end{split}
\end{equation}
where $\bm{\mu} = \displaystyle\sum_{i}{n_i\bm{\hat r_i}}$ and $\bm{\nu} = \displaystyle\sum_{i}{\bm{\hat S_i}}$\\. 
Here $\bm{E} = \bm{E_0}e^{-i \omega t}$ and $\bm{B} = \bm{B_0}e^{-i \omega t}$
are the electric and magnetic fields of the incident light of frequency $\omega$.\\
Using perturbation theory, we calculate the nonlinear optical coefficients
following Orr and Ward\cite{16}. The 2{\ti{nd}} order correction to the 
polarization is given by \cite{b} \\
\begin{equation}\label{2nd_corr}
\begin{split}
&a^{(2)}_n (t)= \\
&{\footnotesize{
		\frac{1}{\hbar^2}\displaystyle\sum_{m} 
		{\frac{\left(\bm{ \hat \mu_{nm}}.\bm{E}(\omega) + \bm{\hat \nu_{nm}}.\bm{B}(\omega) \right)\left(\bm{ \hat \mu_{mg}}.\bm{E}(\omega) + \bm{\hat \nu_{mg}}.\bm{B}(\omega) \right)}{(\omega_{ng}-2\omega) (\omega_{mg}-\omega)} } } } \\
&\hspace{4cm}
\times e^{i(\omega_{mg}-2\omega)t}
\end{split}
\end{equation}
where $\bm{\mu_{ml}} = \langle \phi_m \vert \bm{ \hat \mu} 
\vert \phi_l \rangle$ is the electric dipole moment and 
$\bm{\nu_{ml}} = \langle \phi_m \vert \bm{ \hat S} \vert \phi_l \rangle$ is the
magnetic transition dipole moment.  $\bm{ \hat S} = S_x\bm{\hat x} + 
S_y\bm{\hat y}+ S_z\bm{\hat z}$, $S_i$s are spin matrices. \\
Clearly, the numerator of Eqn.(5) contains terms of the form 
$(\bm{\mu .E})(\bm{\nu. B})$ which leads to the coupling of the electric and 
the magnetic fields.  We calculate second order correction to the polarization
due to electric field and magnetic field by obtaining the expectation values,
\begin{equation}
\bm{\tilde P} =\frac{1}{N} \langle \psi^{(2)} \vert 
\bm{ \hat \mu} \vert \psi^{(2)} \rangle \: \: 
\: \: \bm{\tilde {Ms}} = \frac{1}{N}\langle \psi^{(2)} 
\vert \bm{ \hat \nu} \vert \psi^{(2)} \rangle\: .
\end{equation}
Here $\psi^{(2)}$ is the second order corrected wave function. 
From Eq.\ref{eq:polar} and Eq.\ref{2nd_corr} we get the second order ME 
coefficients as follows,
\begin{align}\label{eq:chi_2}
	&\chi^{(2)}_{ijk} (2\omega,\omega)
	\mspace{0mu}
	\notag\\
	&= \frac{N}{\hbar^2}\displaystyle\sum_{mn}^{\ }\frac{\mu_{gn}^i \mu_{nm}^j \nu_{mg}^k}
	{(\omega_{ng}-2\omega+i\eta)(\omega_{mg}-\omega+i\eta)}
	\notag\\
	&+ \qquad \frac{\mu_{gn}^j \mu_{nm}^i \nu_{mg}^k  }
	{(\omega_{ng}^{*}+\omega)(\omega_{mg}-\omega+i\eta)}
	\notag\\
	&+ \qquad \frac{\mu_{gn}^j \mu_{nm}^k \nu_{mg}^i} 
	{(\omega_{ng}^{*}+\omega)(\omega_{mg}^{*}+2\omega)}
\end{align}
To avoid the value of $\chi^{(2)}_{ijk}$ shooting to very large values near the poles we have added the term $i\eta$ in the denominator. In general, when there are two different input frequencies $\omega_1$ and $\omega_2$, there is an intrinsic permutation symmetry $\mathcal{P}_1$ and $\chi^{(2)}_{ijk}$ should be averaged over all such permutations.
This expression is similar to that obtained in Ref.\cite{a} with the numerator 
consisting of product of both electric and magnetic transition dipole moments.
\begin{equation*}
\mu_{gn}^i \mu_{nm}^j \nu_{mg}^k \ \equiv \langle g 
\vert \hat{\mu}^i \vert n \rangle \langle n \vert \hat{\mu}^j 
\vert m \rangle \langle m \vert \hat{\nu}^k \vert g \rangle
\end{equation*}
Where $\vert g \rangle$ is the ground state and $\vert n \rangle$ and $\vert n \rangle$ are two different excited states of the system.
In the absence of spin-orbit coupling, the Hamiltonian commutes with 
$\hat{S}_z$. So two states connected by the electric dipole moment operator 
$\hat{\mu}_e$(i.e. having different parity) cannot be connected 
by the magnetic dipole operator $\hat{\mu}_b$. Hence the second order ME 
coefficient, $\chi^{(2)}$, would be zero. Only when spin SU(2) symmetry is broken, 
all the states $ \vert g \rangle$, $\vert m \rangle$ and $\vert n \rangle$ 
are no longer eigenstates of the $\hat{S}_z$. So coupling of the transition 
dipole moments could lead to a non-zero $\chi^{(2)}$. Note that in the 
expression for $\tilde{P}^{(2)}$ there are contributions from nonlinear 
optical susceptibilities $\tilde{\chi}^{(2)}_{eee}$ arising only due to the 
electric field of the light which have very large values compared to the 
cross terms, $\tilde{\chi}^{(2)}_{eem}$ or $\tilde{\chi}^{(2)}_{emm}$. \\
In the presence of spin-orbit coupling, the Zeeman perturbation term in Eqn. \ref{eq:perturb} would be
\begin{equation}
	\displaystyle\sum_{i}{\bm{B}.(\bm{\hat{L}_i}+\bm{2\hat{S}_i})} = \sum_{i}{\bm{B}}.(\bm{\hat{J}_i} + \bm{\hat{S}_i})
\end{equation}
rather than $\sum_{i}{\bm{B}}.\bm{\hat{S}_i}$. Here $\bm{\hat{J}}$ is the total angular momentum quantum number. But, since it is a good quantum number in this process, the contribution to $\chi^{(2)}$ due to this $\bm{\hat{J}}$ would be zero, and thus effectively the perturbation term is only due to $\bm{\hat{S}}$. The term could also be written as $\sum_{i}{\bm{B}}.(2\bm{\hat{J}_i} - \bm{\hat{L}_i})$ and the contribution would be due to $\sum_{i}{\bm{B}}.\bm{\hat{L}_i}$. But the results would not change. Since $\bm{\hat{L}_i}$ and $\bm{\hat{S}_i}$ are coupled, taking the one with the smallest number of eigenstates would suffice.
\vspace{-0.5cm}
\section{THE MODEL}
\begin{figure}[h!]
	\centering
	\includegraphics[scale=0.3]{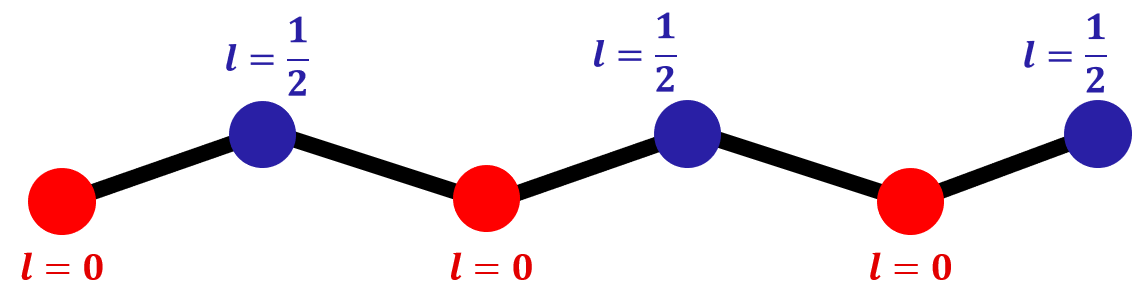}
	\caption{Schematic structure of the model system having alternating sites
 $A$ and $B$.}
	\label{fig:picture}
\end{figure}

We consider a one dimensional (zigzag) chain having alternate 
sites with spin-orbit coupling (FIG.\ref{fig:picture}). The zigzag nature of 
the system ensures that it can be exposed to an extra dimension of electric 
field.  For simplicity,  we have considered only the z-component, $L_z$ of
the orbital angular momentum. $\vert L=\pm\frac{1}{2}\rangle$ are the two 
eigenstates of the z-component of the orbital angular momentum operator 
$\hat{L}_z$ with quantum number  $l =\frac{1}{2}$. The single orbital Hubbard 
model for fermions has 4 degrees of freedom per site, which can be represented 
by $\vert c \rangle$. Hence a fermionic site with a spin-orbit coupling will 
have 8 possibilities, $\vert c \rangle \bigotimes \vert 
L=\pm\frac{1}{2}\rangle $. We have considered the unperturbed Hamiltonian as

\begin{multline}
\hat{H}_0 = \sum_{\langle ij \rangle, \alpha, \sigma} 
t_{ij} d^{\dagger}_{i\alpha \sigma} d_{j \alpha \sigma} + h.c.
+ U \sum_{i} {n_{i\alpha \uparrow}}{n_{i\alpha \downarrow}} \\
 + \lambda\sum_{i}\vec{L}_i. \vec{S}_i
\end{multline}

where $d_{i\sigma} = (c_{i\sigma}\bigotimes\hat{L}^z)_{\alpha\alpha}$. 
$\alpha$ is the pseudospin index for two eigenstates of $L_z$ ($+1/2$ and $-1/2$) and 
$\sigma$ is the index for the spin. Here $n_{i\alpha}=
c^{\dagger}_{i\alpha}c_{i\alpha}$ is the number operator. $t_{ij}$ and 
$U$ are respectively the hopping and Hubbard parameters. $\lambda$ is the strength of spin-orbit coupilng. The term
$\sum_{i}\bm{L_i}. \bm{S_i} = \sum_{i}L^z_iS^z_i + L^+_iS^-_i + L^-_iS^+_i$ is 
considered explicitly to break the SU(2) symmetry.
The last term of the Hamiltonian can also be used for spin-hardcore-boson 
coupling after the  transformation
\begin{equation}\label{eq:holstein}
\begin{split}
&L^+ = \hat{b}^\dagger \\
&L^- = \hat{b} \\
&L^z = \hat{b}^\dagger\hat{b} - \frac{1}{2}
\end{split}
\end{equation}
where $\hat{b}^\dagger$ and $\hat{b}$ be the bosonic creation and annihilation 
operators and $[\hat{b}^{\dagger},\hat{b}] = 1$.Here we have assumed that there is a hardcore boson on each site (either $1$ boson or $0$ boson). These bosons can be coupled with spins which break the SU(2) symmetry.
\section{RESULTS AND DISCUSSIONS}
\begin{figure}[h!]
	\centering{ 
		\includegraphics[width=0.35\textheight,height=0.2\textheight]{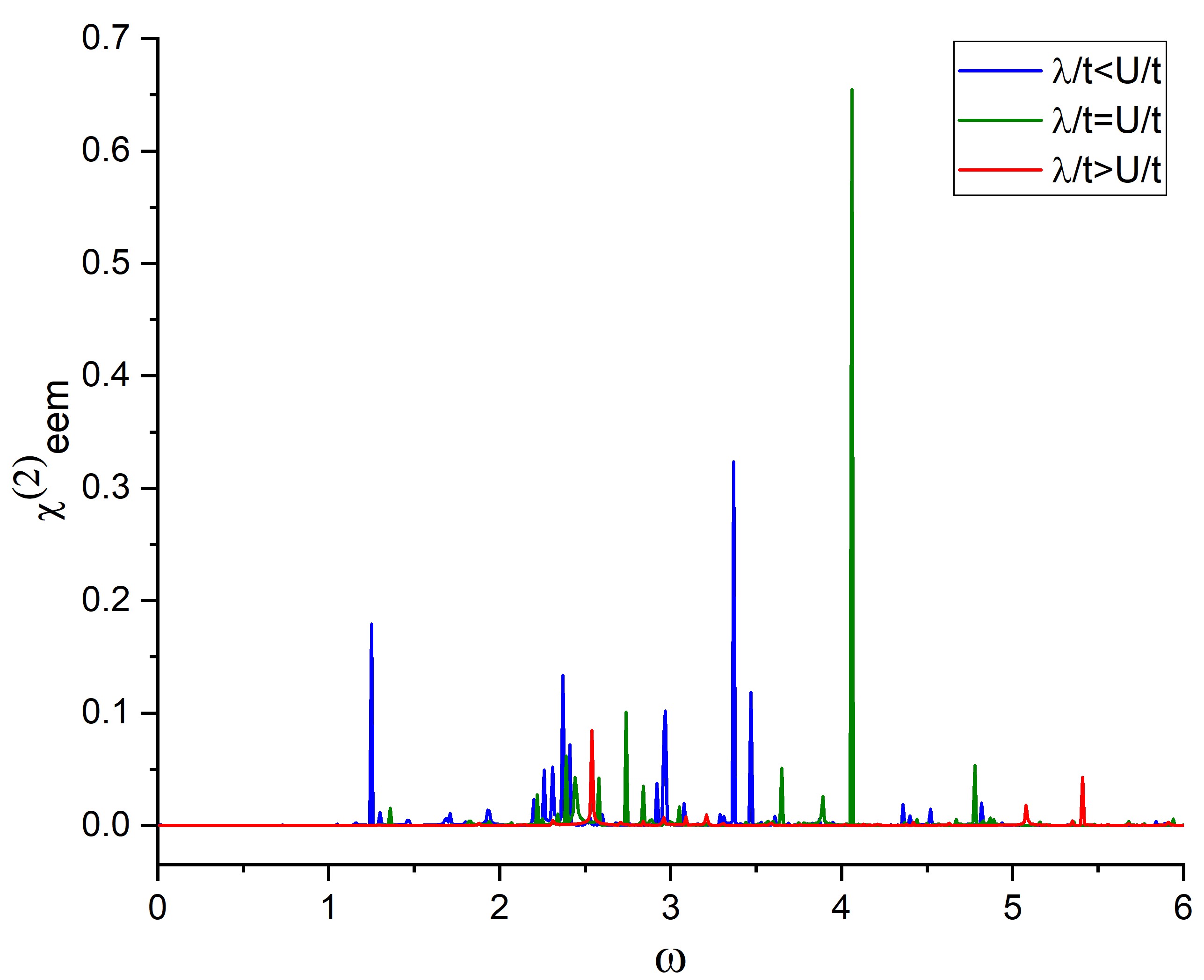}
		\caption{Combined plot of $\chi^{(2)}_{eem}$ vs $\omega$(in units of $\hbar eV$) for $U/t=4.0$ and $\lambda/t = 2.0, 4.0$ and $6.0$. For $\lambda = 2.0 and 4.0$, there are more number of less intense peaks. For $\lambda/t=U/t=4.0$, there is a single peak with high amplitude due to high degeneracy of the excited states involving excitation corresponding to both exchange and orbital transitions.}
		\label{fig:bee_vs_lambda_all}
	}
\end{figure}

Using the above mentioned 
basis set, we have diagonalized  the Hamiltonian matrix exactly for a chain of 6 sites. We have obtained the 200 lowest eigenstates using Lanczos method \cite{Lanczos:1950zz}. Using these states, ME susceptibilities are computed as given in Eq. 6.\\
The tumbling average of these susceptibilities are computed from the expression \cite{tumbling_avg,tumbling_avg2}
\begin{equation}
	\vert\vert \chi^{(2)} \vert\vert = \frac{1}{3} 
	\sqrt{\sum_{i}\sum_{j}\vert \chi^{(2)}_{ijj}+\chi^{(2)}_{jji}+\chi^{(2)}_{jij} \vert^2 }
\end{equation}
These are  experimentally important quantities and also define a scalar 
quantity for better analysis.
In the Hamiltonian we have two parameters, on-site correlation $U/t$ and the SOC strength $\lambda/t$. Now, without these two terms, the solution is a plane wave with delocalized eigenstates. When U = 0, the solution is still a charge delocalized state. On the other hand, when lambda 0 with finite U, the solution is a localized state. thus we ask, how the coefficients would vary in the regimes (i)$U\rightarrow 0, (i)U>\lambda$, (iii) $U\simeq\lambda$ and (iv)$U<\lambda$.\\
\begin{figure}[h!]
	\centering{ 
		\subfloat[]{\includegraphics[width=0.35\textheight,height=0.2
			\textheight]{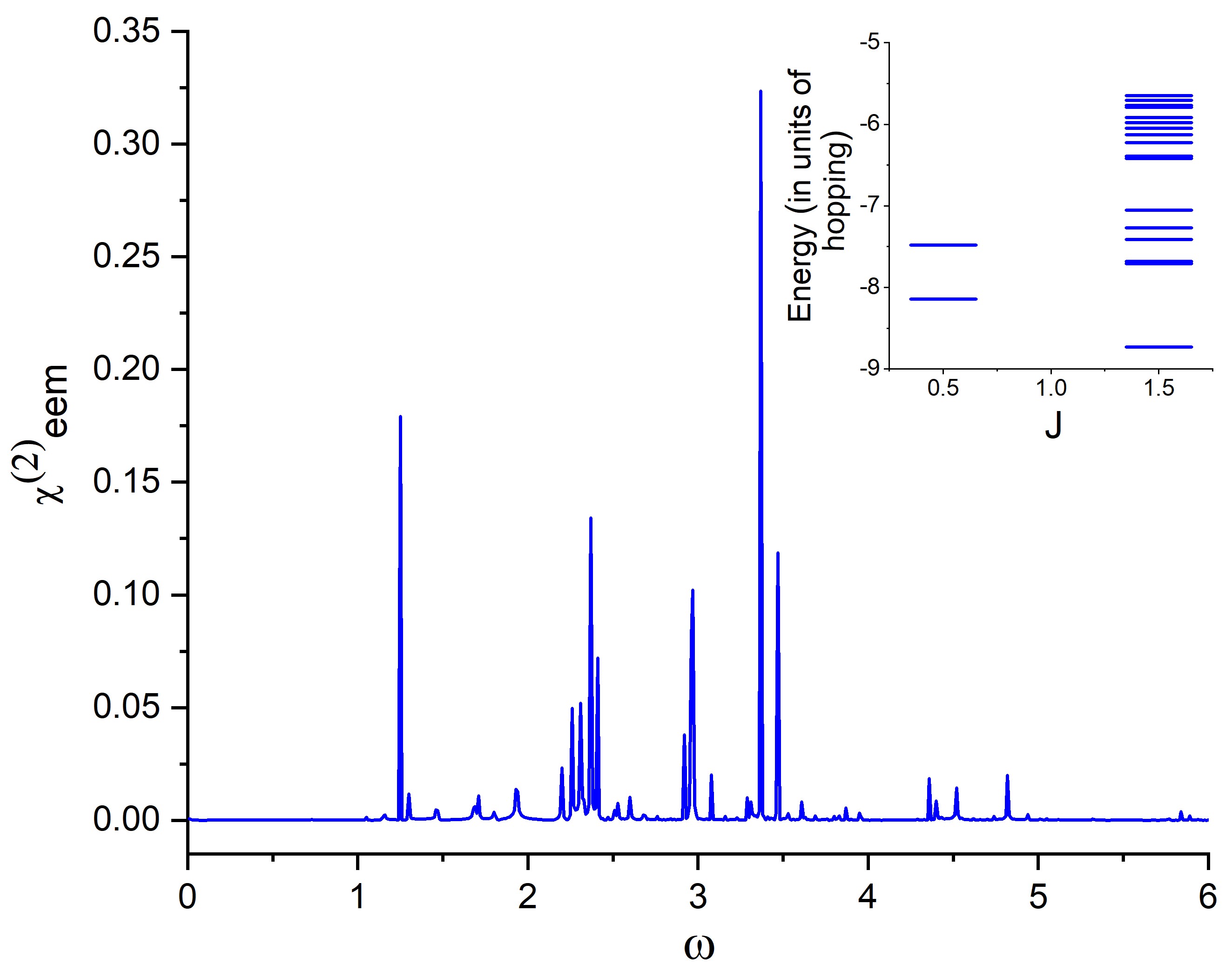}}\\
		\subfloat[]{\includegraphics[width=0.35\textheight,height=0.2
			\textheight]{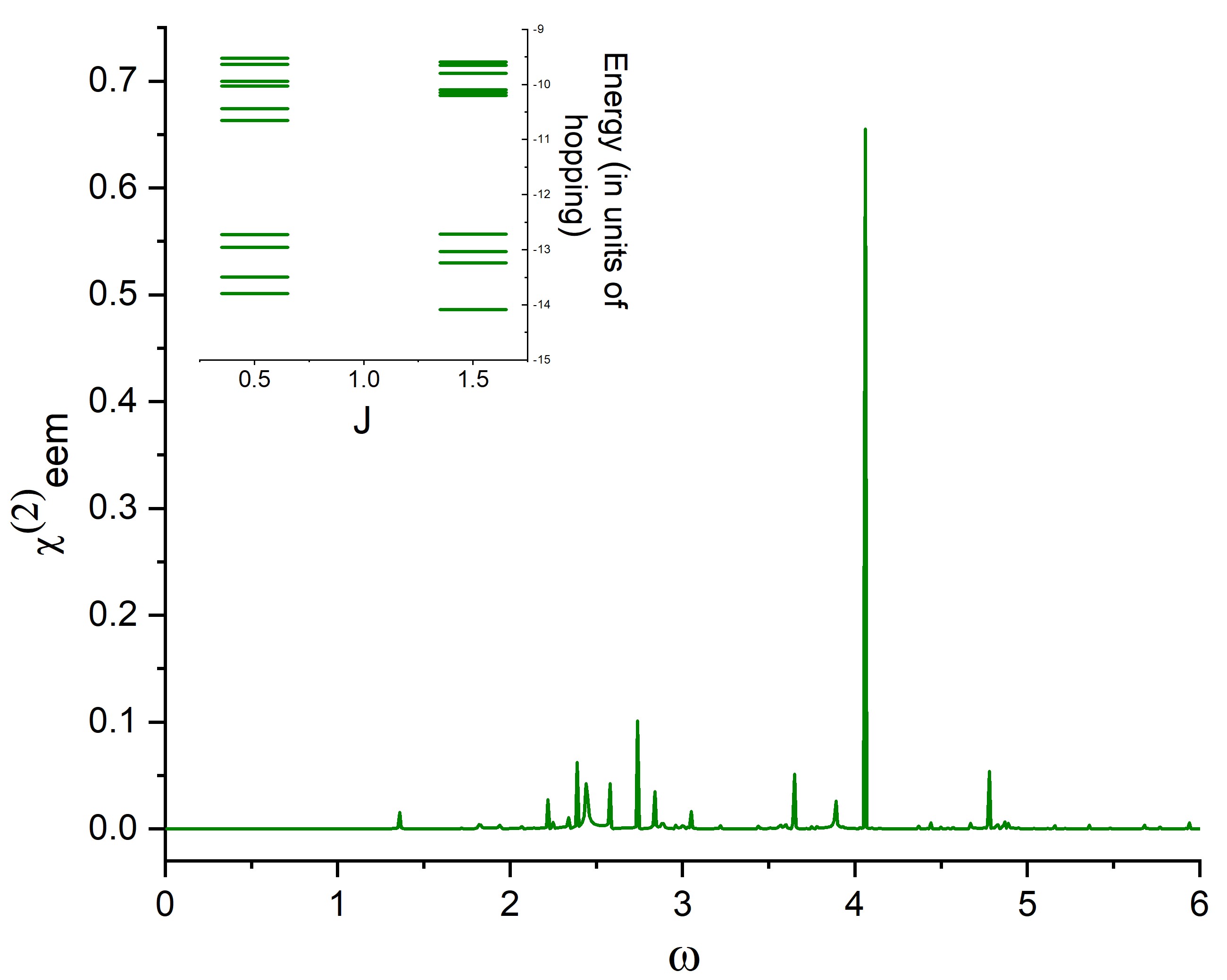}}\\
		\subfloat[]{\includegraphics[width=0.35\textheight,height=0.2\textheight]{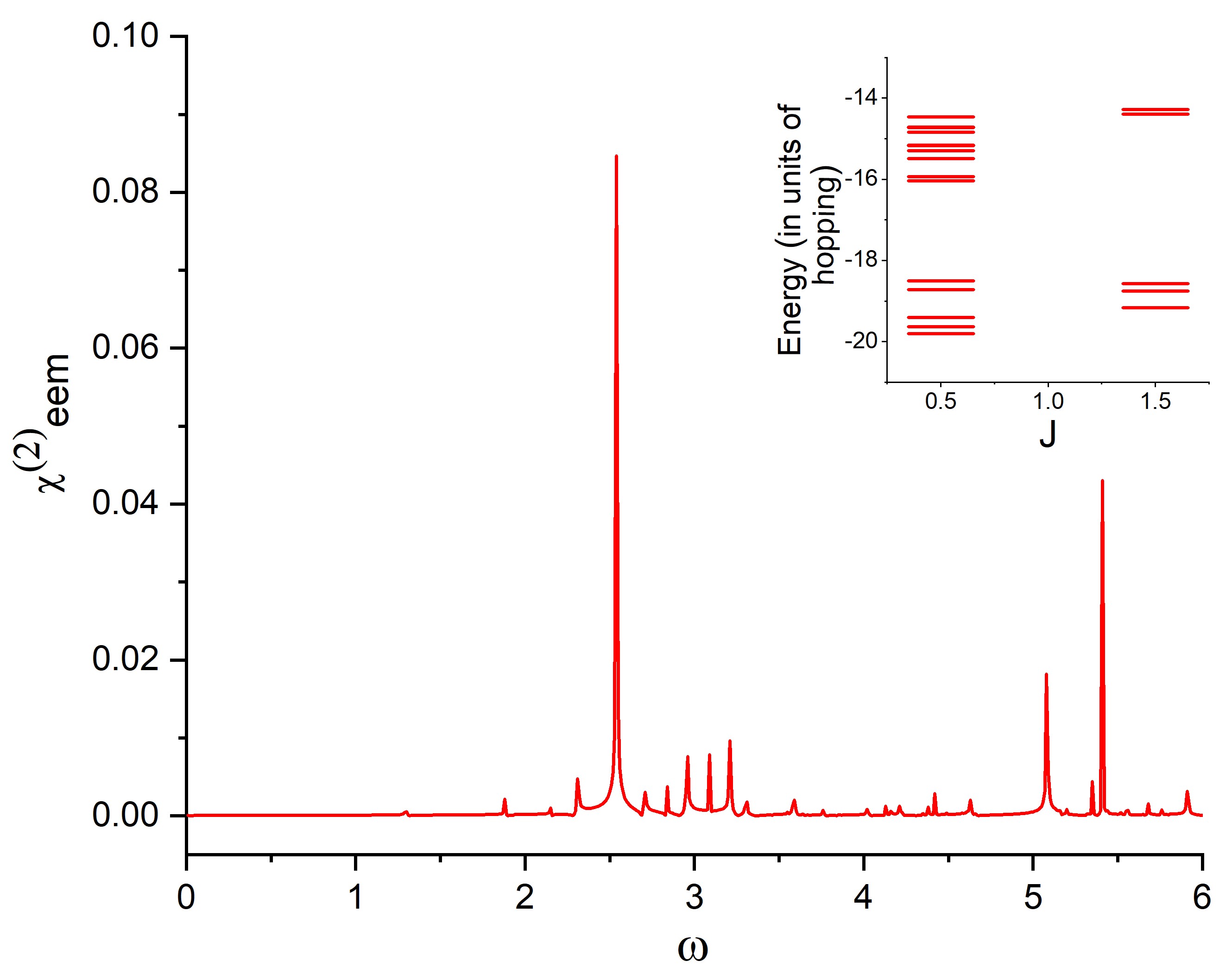}}
		\caption{Plots of $\chi^{(2)}_{eem}$ vs $\omega$ (in units of $\hbar$ eV)for $U/t=4.0$ and $\lambda/t = 2.0, 4.0$ and $6.0$. The insets show plots of energy in eV vs total angular momentum quantum number J in each case. For $\lambda/t = 4.0$, there are equal number of states for both $J=0.5$ and $J=1.5$ and thus there are more degeneracies.}
		\label{fig:bee_vs_lambda}
	}
\end{figure}
For a fixed value of $U/t = 4.0$, we have shown in FIG.\ref{fig:bee_vs_lambda}, the variation of 
$\chi^{(2)}_{eem}$ versus $\omega$ (in units of $\hbar$ eV) for three different values of spin-orbit coupling strength $\lambda/t$ . The plots are obtained by varying $\omega$ in steps of $0.01$ upto a value $6.0$. 
Very large value of $\lambda$ is unphysical for real materials but physically 
realizable in systems of ultracold atoms in optical lattices where the parameters 
can be tuned experimentally.
When $\lambda/t < U/t $, the low energy physics is governed by the strength of $\lambda$. The lowest excitations are those which involve transition from $l_z=\frac{1}{2}$ to $l_z=-\frac{1}{2}$ or vice versa. The excitations corresponding to the exchange involving Hubbard U will have higher energy. The opposite is the case in the $\lambda/t > U/t $ regime, in which the energy is lower for excitations due to magnetic exchange than those due to change in orbital angular momentum. For $\lambda/t \simeq U/t $, the energies corresponding to both the excitations are similar, that is, the excited states are highly degenerate. This is obvious from FIG. \ref{fig:bee_vs_lambda_all}, in which all of FIG.\ref{fig:bee_vs_lambda} (a), (b) and (c) are superposed. There are some less or moderately intense peaks for $\lambda/t < U/t $ and $\lambda/t > U/t $ due to less degeneracy of the excited states. And, for $\lambda/t \simeq U/t$,  there are less number of peaks, but a single peak with high amplitude corresponding to high degeneracy of the excited states.
The insets of FIG.\ref{fig:bee_vs_lambda}(a), (b) and (c) also verify the above result. Here we have plotted the
energies of 20 lowest excited states as a function of total angular momentum $J$, computed from the expectation value of $\hat{J}^2$ operator($\langle \hat{J}^2 \rangle = \langle (\hat{L}+\hat{S})^2\rangle$).It is evident from these insets that, for $\lambda/t<U/t$, the low-energy excited states will have higher $J$ values as the electrons prefer filling different angular momentum states rather than pairing up in one orbital, hence the density of states is high at $J=1.5$. For $\lambda/t>U/t$ the, exchange is preferred, hence the states with lower value of $J$ , namely $J = 0.5$ have more population of states. At $\lambda/t = U/t$, the excited states comprise excitations due both exchange and spin-orbit coupling and hence the are almost equal number of states having $J = 0.5$ and $J = 1.5$.\\
\begin{figure}[h!]
	\includegraphics[width=0.37\textheight,height=0.50\textheight]{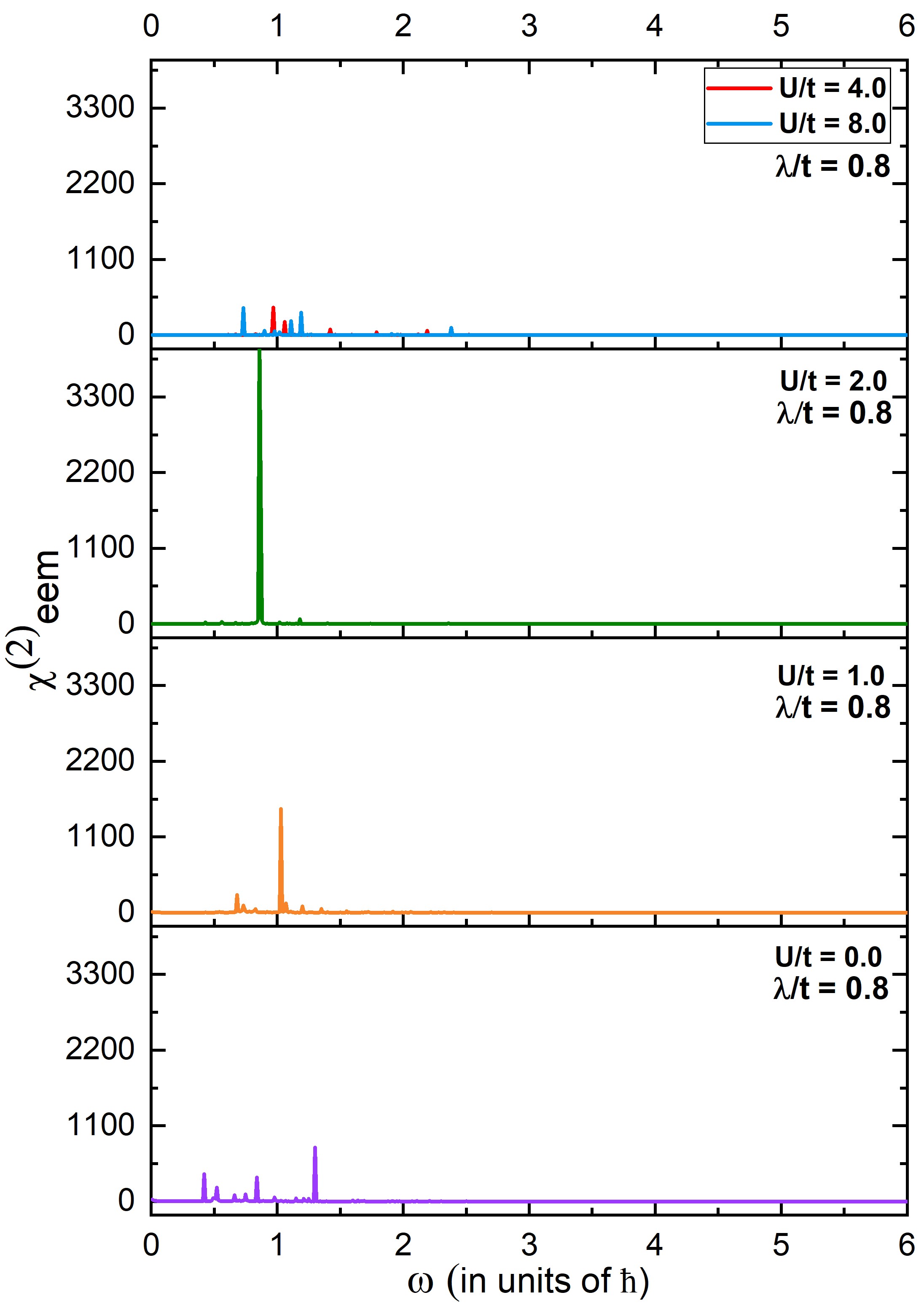}
	\caption{Plots of $\chi^{(2)}_{eem}$ vs. $\omega$ for different values 
		of $U/t = 0.0, 1.0, 2.0, 4.0, 8.0$. For $U=1.0$ and $2.0$, the peaks have greater amplitude owing to similar excitation energies due to SOC and exchange. }
	\label{fig:bee_vs_U}
\end{figure}
 The plots of  $\chi^{(2)}_{eem}$ versus $\omega$ for five different values of $U/t$ for a fixed $\lambda/t= 0.8$ , are shown in FIG.\ref{fig:bee_vs_U},
The spin excitations cost energy $\approx \lambda$ and at most $3\lambda$ since there are 3 sites with SOC in our system. So we find a high amplitude peak in the regime $U/t= 0.8 - 2.4$ as excitations due to $U$ and $\lambda$ have similar energies, leading to more degeneracies. For $U<\lambda$ and $U>3\lambda$ the degeneracies are broken and so the amplitude of the peaks decrease. Also, for higher $U$ values, very few peaks are visible only at lower values of $\omega$, namely, those connecting states having similar $\langle U\rangle$. This is because the spin excitations together cannot match the excitations costing energy $U$ and so peaks at higher $\omega$ values are not probable.\\
\indent From FIG. \ref{fig:bee_vs_U}, it is also evident that there are non-zero ME susceptibilities at $U=0$. In this regime, the model is effectively tight-binding with spin-orbit coupling. Though the electronic spins are completely delocalized in this case so there is no role of kinetic exchange. But the broken SU(2) symmetry in the presence of spin-orbit coupling leads to spin-orbital excitations between different eigenstates and hence it is sufficient to give nonzero $\chi^{(2)}_{eem}(\omega)$.\\
 The effect of SU(2) symmetry breaking upon spin-phonon coupling 
\cite{sph1,sph2,sph3} will be similar when we consider hardcore bosons, owing to the equivalence of the two, as shown in Eqn.(\ref{eq:holstein}).
But in reality, there are many bosonic modes, so one has to use Holstein-Primakoff 
transformations \cite{holsteinpaper} to obtain $\bm{\hat{L}}$ operator from the bosonic operators and thus the Hamiltonian for spin-orbit 
coupling can also be used. In that case, there will be many eigenstates for $\hat{L}_z$. So, for low values of $\lambda$ there will be many possibilities for transitions among $l$ values giving rise to many more peaks.
%\vspace{-1.0cm}
\section{CONCLUSION}
In conclusion, we have shown that there can be non-zero second order dynamical ME susceptibility $\chi^{(2)}_{eem}(\omega)$ at certain resonant frequencies in a system when the spin SU(2) symmetry is broken by spin-orbit or spin-phonon coupling. These resonant frequencies correspond to the different spin and charge excitations in case pf spin-orbit coupling. For spin-phonon coupling, these correspond to charge excitations as well as excitations between different phonon modes. The amplitude of the peaks are very high when both the excitations are in similar energy range.
\begin{acknowledgments}
A.L. is grateful for the financial support from DST and CSIR of the Government of India.
\end{acknowledgments}
\bibliographystyle{aipauth4-1}
\bibliography{me}
\end{document}